\documentclass[twocolumn,reprint]{revtex4-1}

\usepackage[english]{babel}
\usepackage{graphicx}
\usepackage{epstopdf}
\usepackage{amssymb}
\usepackage{amsmath}

\begin{document}


\title{Obtaining imaginary weak values with a classical apparatus: applications for the time and frequency domains.}

\author{Yaron Kedem }
\email{kedem@kth.se}
\affiliation{Raymond and Beverly Sackler School of Physics and Astronomy, Tel-Aviv University, Tel-Aviv 69978, Israel}
\affiliation{ Nordic Institute for Theoretical Physics (NORDITA), Roslagstullsbacken 23, S-106 91 Stockholm, Sweden}

\begin{abstract}
Weak measurements with imaginary weak values are reexamined in light of recent experimental results. The shift of the meter, due to the imaginary part of the weak value, is derived via the probability of postselection, which allows considering the meter as a distribution of a classical variable. The derivation results in a simple relation between the change in the distribution and its variance. By applying this relation to several experimental results, in which the meter involved the time and frequency domains, it is shown to be especially suitable for scenarios of that kind. The practical and conceptual implications of a measurement method, which is based on this relation, are discussed. 
\end{abstract}


\maketitle

\section{Introduction}

Weak values were introduced, in a seminal paper by Aharonov, Albert and Vaidman \cite{AAV88}, as the result of a weak measurement on a pre- and postselected system. Their ideas were met with some suspicion \cite{peres, leget} but have been shown to be valid \cite{av}. Since then, they were used for various tasks such as directly measuring quantum states \cite{lund, traj, pol} or observation of tiny effects \cite{HK,How}. Even though their practical benefits were questioned \cite{knee}, many new schemes for utilizing weak values are being published rather frequently in recent years. A wide range of challenges, such as charge detection \cite{charge}, measuring small time delays \cite{Brunner,phase,Struebi} or observing Kerr nonlinearity \cite{stein}, were addressed. Additional improvements, such using orbital-angular momentum \cite{angular}, were demonstrated and some extensions to the formalism were suggested \cite{KV,cont}. 

Recently, weak measurements were demonstrated, using the time and frequency domains, in a number of experiments: improving phase estimation \cite{led,sub}, measuring velocity \cite{dopler} and studying atomic spontaneous emission \cite{atom}. In all these schemes, imaginary weak values were used in order to make transformations between effects in time and frequency. In \cite{led,sub} a time delay was converted to a spectral shift and in \cite{dopler,atom} it was vice versa. The treatment of the time and frequency domains as a measurement device (meter) was originally suggested by Brunner and Simon \cite{Brunner} and it is in some contrast to the usual formalism of weak measurement, where the meter is described using quantum variables such as position and momentum. In case one wishes to treat time and frequencies as quantum variables, some conceptual difficulties might be encountered. In this work, we provide a common theoretical framework for the experimental results, which is focused on the measurement of imaginary weak values. It is based on the use of a classical random variable for describing the meter, rather than a wavefunction. The formalism is general for any scenario involving imaginary weak values, and it can be applied for a wide range of weak measurements schemes. 

The weak value of an observable $C$ on a pre and postselected system, described by the {\em two-state vector} $\langle \Phi |~|\Psi\rangle $, is given by
\begin{equation}\label{wv}
C_w \equiv { \langle{\Phi} \vert C \vert\Psi\rangle \over
\langle{\Phi}\vert{\Psi}\rangle } .
\end{equation}
A few properties of this expression differ it from other values that can be assigned to an observable, like an expectation value or eigenvalues. It can be much larger, if the pre and post-selection states are nearly orthogonal, and it is complex in general \cite{Joz}. The imaginary part of the weak value was found to be highly useful for practical goals \cite{snr} and its significance was broadly discussed \cite{sig}. Imaginary weak values were used in most, if not all, of the experiments showing increased precision. 

\section{The standard formalism}

The standard formalism of weak measurements is based on an interaction between a pre- and postselected system to a meter, which is also considered as a quantum system. The interaction can be represented using a Hamiltonian 
\begin{equation}\label{h1}
H = g(t) PC ,
\end{equation}
where $C$ is an observable on the system, $P$ is an operator on the meter and $g(t)$ is a coupling function satisfying $ \int g(t) dt =k$. If the strength of this interaction is small, the wavefunction of the meter is real valued and the system is pre- and postselected to $\langle \Phi |~|\Psi\rangle $, the change in the average of $Q$, a variable conjugate to $P$, would be \cite{AV90}
\begin{equation}\label{changeQ}
\delta Q = k {\text Re}C_w,
\end{equation}
and the change in the average of $P$ would be 
\begin{equation}\label{changeP}
\delta P =2 k {\text Im}C_w \text{Var}\left( P \right),
\end{equation}
where $\text{Var}\left( P \right) = \langle P^2 \rangle - \langle P \rangle^2 $ is the variance of $P$. Here, we can consider the average $\left\langle \bullet \right\rangle$ to be taken with respect to the initial wavefunction of the meter. Later, we will extend the notion of average to encompass a more statistical distribution. 

The shifts (\ref{changeQ}) and (\ref{changeP}) can be derived using the AAV effect, i.e. replacing the operator $C$ in (\ref{h1}) by its weak value and calculating the evolution of the meter under the effective Hamiltonian. In the case $C_w$ is real this Hamiltonian is self-adjoint, which corresponds to a unitary evolution. When $C_w$ is complex, the resulting non unitary evolution of the meter might seem unphysical, especially since $P$ is a constant of motion under the Hamiltonian (\ref{h1}). Below, we will offer an alternative derivation of (\ref{changeP}) and show that unlike (\ref{changeQ}) it does not require interference in the wavefunction of the meter.

\section{Derivation of the main result}

Let us consider a simpler Hamiltonian
\begin{equation}\label{h2}
H = \tilde{g}(t) C ,
\end{equation}
where $\tilde{g} (t)$ is a coupling function satisfying $ \int \tilde{g} (t) dt =\tilde{k}$. With the assignments $\tilde{g}(t) = g(t) P$ and $\tilde{k} = k P$, we can recover the interaction (\ref{h1}), but we can also regard $\tilde{k}$ as a parameter so (\ref{h2}) would operate only on the Hilbert space of the system. Since our interest is in the regime of weak interactions we can assume $\tilde{k} \ll 1$. If the system is initially in a state $|\Psi\rangle $, then after the evolution caused by (\ref{h2}), the probability of finding it in a state $|\Phi\rangle $, for a known $\tilde{k}$ is given by 
\begin{eqnarray}\label{prob}
P\left(| \Phi\rangle \Big|\tilde{k}\right) &=& \left| \langle \Phi | e^{-i \tilde{k} C} | \Psi\rangle \right| ^2 \nonumber\\
&=& \left| \langle \Phi | \Psi\rangle \right| ^2 \left( 1+ 2 \tilde{k} {\text Im} C_w \right) + O(\tilde{k}^2) .
\end{eqnarray}

Now let us consider a situation where the value of $\tilde{k}$ varies according to some distribution $f \left(\tilde{k} \right) $. This is to say that the experiment is repeated many times and in each run $\tilde{k}$ can obtain a different value, where the probability that $\tilde{k} =x$ is $f(x)$ if $\tilde{k}$ is discrete or $f(x) dx$ if it is continuous. Using this distribution we can calculate different moments of $\tilde{k}$, for example its average is given by $\left\langle \tilde{k}\right\rangle = \int \tilde{k}f \left(\tilde{k} \right) d\tilde{k}$. For the interaction to be weak, $ f \left(\tilde{k} \right) $ should have a significant value only where $|\tilde{k}| \ll 1$ so the average of $\tilde{k}$ for this distribution, or any of its moments, should be small. We can later relax this requirement to have only the width of the distribution small.

A post-selection to $|\Phi\rangle $ means we are interested only in the cases where the system was found in the state $|\Phi\rangle $. Since the probability for this depends on $\tilde{k}$, the post-selection will modify the distribution of $\tilde{k}$. According to Bayes' theorem, the probability to get some value of $\tilde{k}$, given a post-selection $|\Phi\rangle $, is 
\begin{equation}\label{dist}
f_{ \Phi } \left(\tilde{k} \right) = { f \left(\tilde{k} \right) P\left (|\Phi\rangle \Big| \tilde{k} \right) \over P \left( |\Phi\rangle \right) }
\end{equation}
where $ P\left( |\Phi\rangle\right) = \int P\left(| \Phi\rangle \Big|\tilde{k}\right) f \left(\tilde{k} \right) d\tilde{k} \simeq \left| \langle \Phi | \Psi\rangle \right| ^2 \left( 1+ 2 \left\langle \tilde{k}\right\rangle {\text Im} C_w \right) $ is the average probability of post-selection. By inserting (\ref{prob}) into (\ref{dist}) we can calculate the modified average of $\tilde{k}$, up to second order in $\tilde{k}$:
\begin{eqnarray}\label{modk}
\left\langle \tilde{k}\right\rangle _{ \Phi} &=& \int \tilde{k}f_{ \Phi } \left(\tilde{k} \right) d\tilde{k} \nonumber\\ 
&\simeq& {\int \tilde{k}\left( 1+ 2 \tilde{k} {\text Im} C_w \right) f \left(\tilde{k} \right) d\tilde{k} \over 1+ 2 \left\langle \tilde{k}\right\rangle {\text Im} C_w} \nonumber\\
&\simeq& \left\langle\tilde{k}\right\rangle + 2 {\text Im} C_w \left( \left\langle \tilde{k}^2\right\rangle - \left\langle \tilde{k}\right\rangle^2 \right).
\end{eqnarray}
A quantity of interest for observing some effect in an experiment can be the difference between the postselected and initial averages
\begin{equation}\label{delk}
\delta \tilde{k}= \left\langle \tilde{k}\right\rangle _{ \Phi} - \left\langle \tilde{k}\right\rangle \simeq 2 {\text Im} C_w \text{Var}\left( \tilde{k}\right),
\end{equation}
where $\text{Var}\left( \tilde{k}\right)= \left\langle \tilde{k}^2\right\rangle - \left\langle \tilde{k}\right\rangle^2 = \left( \Delta \tilde{k}\right)^2$ is the initial variance of $\tilde{k}$. This simple relation between the change in a parameter and its uncertainty is our main result.
It should be noted that this result does not depend on the specific form of $ f \left(\tilde{k} \right)$, i.e. it is not assumed to be, for example, Gaussian. The only assumption, which leads to the absence of higher order terms in the result, is that $ f \left(\tilde{k} \right)$ have significant values only where $\tilde{k}$ is small. This assumption is discussed in details in sec \ref{weakcond}.

We can see that if $\tilde{k} = k P$, where $k$ is constant and only $P$ varies, the result (\ref{delk}) is the same as (\ref{changeP}). The alternative derivation highlights the fact that the variance appearing there is valid for any type of variations, and not only to pure quantum uncertainty. Naturally, quantum mechanics provides a complete description of any system, so one can argue that any variation in the value of a physical quantity is essentially quantum uncertainty. However, considering a fully quantum description can unnecessarily complicate the analysis of an experimental setup. A formalism involving the distribution of a classical parameter can be much simpler than a complete quantum description.

\subsection{The regime for the validity of weakness} \label{weakcond}

The result (\ref{delk}) regards only the change and variance of $\tilde {k} $ and thus it is independent of its average. That is to say, if we add some known constant to $\tilde {k} $, the difference between the initial and post selected averages will not be affected, as long as we stay in the regime where $|\tilde {k}| \ll 1$. As we will now show, the result (\ref{delk}) can hold even when $\left\langle \tilde{k}\right\rangle$ is not negligible, provided that we take it into account by modifying $C_w$. By doing this, we can treat separately the known part of $\tilde {k} $, which is its average $\left\langle \tilde{k}\right\rangle$, and the unknown part, which is represented by its uncertainty $\Delta \tilde{k}$. 

The evolution $U= e^{-i \tilde{k} C}$, caused by (\ref{h2}), can be written as $U= U_1 U_2$, where $U_1 = e^{-i \left(\tilde{k}- \left\langle \tilde{k}\right\rangle \right) C}$ and $U_2 = e^{-i \left\langle \tilde{k}\right\rangle C}$. Thus, the probability of postselection is given by $\left| \langle \Phi | U_1 | \Psi'\rangle \right| ^2$, where $| \Psi'\rangle = U_2 | \Psi\rangle$. By repeating the calculations of (\ref{prob}), (\ref{dist}) and (\ref{modk}), we can see that eq. (\ref{delk}) is unchanged except for the weak value itself, which is given by 
$C_w \equiv { \langle{\Phi} \vert C \vert\Psi'\rangle \over
\langle{\Phi}\vert {\Psi'}\rangle }$ . Now, the calculations involved only the deviation $\tilde{k}- \left\langle \tilde{k}\right\rangle $ and all moments higher than 2, of this quantity, have been neglected. Each moment was also multiplied by the real or imaginary parts of an expression of the form $\left(\left( C^n \right)_w\right)^m$ for some $n, m$. Strictly speaking, all these terms have to be small, but in order to see this explicitly, one should specify the distribution $ f \left(\tilde{k} \right)$, the state $\langle \Phi |~|\Psi\rangle $ and the observable $C$. However, in case the second moment, $\text{Var}\left( \tilde{k}\right)$, is large, higher (even) moments cannot be small. Moreover, in order for the weak value expressions to be large, the scalar product in the denominator $ \langle{\Phi}\vert {\Psi'}\rangle $, which appears in all of them, have to be small. Thus, a necessary condition for the validity of (\ref{delk}) is that 
\begin{equation}\label{cond}
{\text Im} C_w \Delta \tilde{k} \ll 1.
\end{equation}
This also implies that the shift $\delta \tilde{k} $ is always smaller than the uncertainty $\Delta \tilde{k}$. 

\section{Discussion}

\subsection{Applying the result for explanation of recent experiments} 

Let us show how this formalism applies straightforwardly to the recent experimental results we have mentioned before. In \cite{led}, a scheme for sensitive phase estimation using white light was implemented. Waveplates were used such that a pulse of light, with one polarization, would reach the output port with a tiny delay $\tau$, compared to a pulse with the orthogonal polarization. The Hamiltonian for photons $H = \omega$ ($\hbar =1$) does not depend on the polarization but the time the Hamiltonian operates is longer for one of the polarization components. This fact can be expressed by considering another Hamiltonian $H = \omega C$, where $C$ is a projection on one polarization, which operates only for time $\tau$. Thus, in this case $\tilde{k} = \omega \tau $ and using eq. (\ref{delk}) we can calculate the change in this parameter. In the experiment, the value of $\tau$ was rather stable, i.e. it did not fluctuate considerably, but the incoming light had a wide spectrum, so $\omega$ had a distribution with large uncertainty. So the change (\ref{delk}) in $\tilde{k}$ is based on a change in $\omega$ which is given by $ \delta \omega \simeq 2 {\text Im} C_w \tau \text{Var}\left(\omega \right) $. Using a spectrometer it is straightforward to detect $\delta \omega $. The simplicity of the result and the fact that it does not depend on experimental details, like the specific shape of the spectrum, allow us to see how a manipulation of the key factors, which are $\text{Var}\left(\omega \right) $ and ${\text Im} C_w $, can yield an efficient estimation of $ \tau$. Furthermore, $ \tau $ can represent any phase difference, so this scheme is applicable to any phase estimation task.

The scheme in \cite{sub} is similar to \cite{led}. The main differences are that the time delay was induced using a Michelson interferometer with one path slightly longer and that a femtosecond fiber laser was used instead of a commercial LED. So the analysis in the previous paragraph is applicable to this experiment as well. 

In \cite{atom}, weak measurements were demonstrated using atomic spontaneous emission. They used an atom with two excited states with magnetic quantum numbers $C=1,-1$. A magnetic field was applied to create a split of $2\Omega$ between the energy levels of the two states so the relevant Hamiltonian can be written as $H= \Omega C$. The atom is excited by absorbing a photon and then spontaneously decays while emitting a photon. The state of the atom right after it absorbs a photon is given by the polarization of the absorbed photon and the polarization of the emitted photon is given by the state of the atom at the time of emission. Since there is no energy difference for the different polarization of the photons the Hamiltonian operates only for a time $t$ when the atom is excited, i.e. between absorption and emission. So for this scheme, $\tilde{k} = \Omega t$. In this experiment, $\Omega $ is rather stable but the time until the decay is exponentially distributed, so the change  (\ref{delk}) is based on a change in $t$ which is given by $ \delta t \simeq 2 {\text Im} C_w \Omega \text{Var}\left(t \right) $. Such a relation can assist in studying the decay process which is highly important in application for quantum computations. 

In \cite{dopler} a technique for measuring velocity, based on weak value, was demonstrated. The setup include a Michelson interferometer where one of the mirrors of was moving in a speed $v$, which changed the length of one path by $vt$, where $t$ is the time when the light hits that mirror. The additional phase accumulated due to the extra segment is given by $\tilde{k} = v t \frac{2 \pi}{\lambda} $, where $\lambda$ is the wavelength on the light. A narrow spectrum was used so $\lambda$ was well defined and the momentary speed of the mirror was rather stable but a pulse with a wide temporal profile was used. So the change (\ref{delk}) was based on a change in the arrival time which is given by $ \delta t \simeq 2 {\text Im} C_w v \frac{2 \pi}{\lambda} \text{Var}\left(t \right) $, where $\text{Var}\left(t \right) $ is the variance of the pulse temporal profile. Here, again, from the simplicity of the result one can see which factors are important and how these factors contribute to the efficiency in which $v$ can be obtained.

The fact that eq. (\ref{delk}) is applicable for schemes of such different nature, testify to its generality. It highlights the two important quantities in any such setup: the weak value of the quantum system and the variance of an experimental parameter. The physical manifestation of these quantities, and how they should be calculated, depends on the details of the specific scheme. Nonetheless, as long as eq (\ref{cond}) is fulfilled, the validity of (\ref{delk}) is independent of those details. This simplicity and clarity can be highly beneficial for devising new schemes by giving an intuition of how changing different parameters will affect the final outcome.

\subsection{Comparison of different measurement schemes}

\begin{figure}
\centering
\includegraphics[trim=2.2cm 0.2cm 7.2cm 0.3cm, clip=true, width=0.5\textwidth]{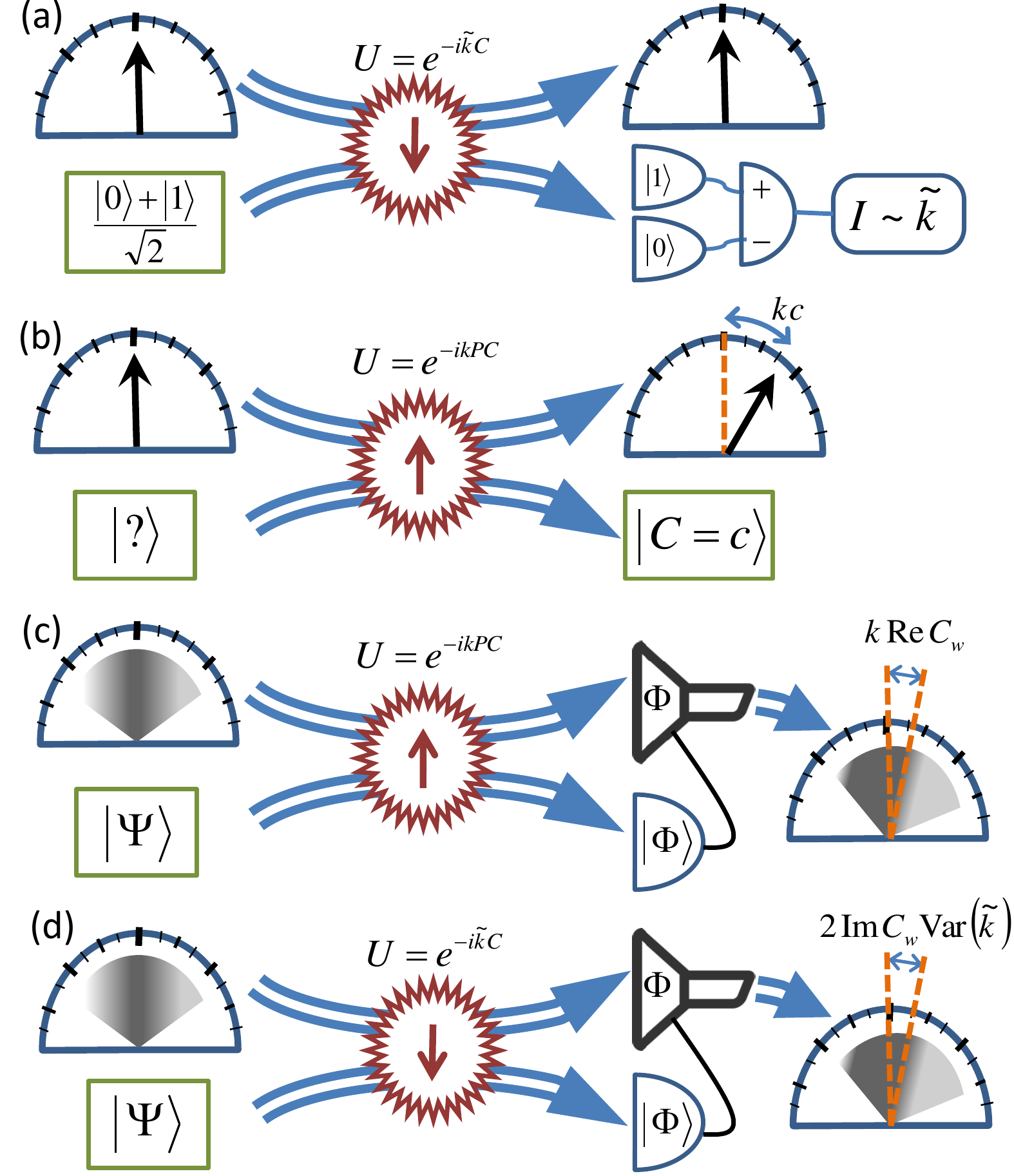}
\caption{ (Color online) An illustration of four measurement schemes: (a) Standard interferometry: All the elements that contribute to the effect are set to as precise a value as possible. Such an element can be regarded as a meter, even though this is typically not done. The system is set in a superposition that creates interference in the output ports, where the effect increases the amplitude in one port and decreases it for another. The strength of the effect can be derived from the difference between the outputs. (b) Von Neumann measurement scheme: The initial state of the system is unknown and the meter is prepared so that its pointer variable has small uncertainty. For any eigenstate of the observable $C$, the interaction shifts the pointer variable by $k c$, where c is the eigenvalue of $C$. (c) Weak measurements using real weak values: The initial state of the system is known and there is large uncertainty in the initial state of the meter. The interaction change the state of the meter into a superposition of shifts and the postselection $\vert {\Phi}\rangle $ creates interference of these shifts, which yield the real part of the weak value. (d) Weak measurements using imaginary weak values: The scheme is similar to (c) but the interaction is changing the state of the system, instead of changing the state of the meter, such that the probability of postselection is dependent on the state of the meter. 
}
\label{fig1}
\end{figure}

The method of weak measurements is an extension of the Von Neumann measurement scheme. While the modifications are quite minor for real weak values, there is a significant difference for imaginary ones. Regarding this method as a practical technique for improving precision, it should also be compared with the standard interferometry scheme. These two schemes, along with schemes for measurement of real and imaginary weak values, are illustrated in Fig \ref{fig1}, showing the similarities and differences with regards to (i) how the system and meter are prepared, (ii) which part is affected by the interaction and (iii) how the information is read out.

The different elements in Fig \ref{fig1} are presented in an abstract way so they can apply to a wide range of scenarios. The system represents any superposition that can produce interference, such as polarizations, different optical paths, spin of a particle etc. The meter refers to any variable that can be monitored. For measurement of real weak values the meter must interfere with itself. This implies that other basis can be used, but the shift (\ref{changeQ}) applies to a specific variable, which we refer to as the meter. For imaginary weak values, the shift (\ref{changeP}) applies to another (conjugate) variable, which can be referred to as the meter, but using (\ref{delk}) we can consider only a single parameter. In standard interferometery one might not identify any meter but many experimental parameters can be considered a meter. An obvious example is the spectrum of light: while most interferometric schemes require a narrow linewidth, a measurement of the spectrum can provide valuable information and might yield significant improvement to the setup. 

From a practical perspective, there can be many considerations for choosing a method and none of the methods can be superior in every possible scenario. Brunner and Simon \cite{Brunner} have shown that measurement of imaginary weak values can outperform standard interferometery when the main limiting factor is alignment errors. For a comprehensive review on the usability of weak value for improving measurements see \cite{dressel}. We will focus on issues regarding the meter and the way it is handled. As mentioned before, this is usually not done explicitly in interferometery and referring to an experimental parameter, which is assumed to be controlled, as a meter might seem unconventional. Nonetheless, in practice, controlling a physical parameter, such as temperature or location of some object, is often done via monitoring its value and correcting if necessary. So treating such a parameter as meter with an average value, which is readout during the experiment, and having some uncertainty, can be very suitable. In such a case, the prior readout can be regarded as a preparation and the final readout might yield valuable information, if the parameter, or meter, was affected by the interaction. When an element in a setup has large uncertainty, the setup is probing a wide range of parameter configurations and a lot of information can be obtained from a final measurement. Eq. (\ref{delk}) offers a straight forward way to extract this information and connect it directly to the quantity of interest. 

Conceptually, weak measurements with real weak value are very similar to a Von Neumann measurement. If one consider the measurement to be repeated many times such that the average shift of the meter amount to $k\left\langle C \right\rangle$, then the weak measurements scheme is simply a modification $\left\langle C \right\rangle \rightarrow {\text Re}C_w$, due to the postselection. For imaginary weak values, the situation is different. In such a scheme, the distribution $f \left(\tilde{k} \right) $ can play the role of a meter and it is unaffected by the interaction itself. On the contrary, the state of the quantum system is changed, according to the value of $\tilde{k} $, an effect that is often called backaction. The meter, i.e. the distribution $f \left(\tilde{k} \right) $, is changed only after the postselection, and only then one can obtain information about the interaction. 

An important component in the approach of the Von Neumann scheme is that the meter can be seen both as a quantum and a classical variable. As we pointed out before, regarding eq. (\ref{delk}), one can obtain exactly the same result by assuming $\tilde{k}$ is a classic or quantum variable (the integration over time can be taken to amount to 1). Besides being a manifestation of the Von Neumann approach, this means that the uncertainty of the meter can have different nature depending on how we describe it. For a quantum system, there is a lower bound, which connects the uncertainties of different bases. Thus, the small uncertainty, which is required for a strong measurement, implies a large uncertainty in the conjugate variable, which is the one used for obtaining the imaginary weak value. Associating this conjugate variable with the parameter $\tilde{k}$ in (\ref{delk}), we can see that we recover the same requirement for uncertainty, since for vanishing uncertainty in $\tilde{k}$ there would be no shift.

\section{Conclusion}

We derived the effect of imaginary weak values using the probability of postselection, through a change in the distribution of a parameter, which functions as a meter. The alternative formalism is applicable to a few recent experimental results and provides a common theoretical basis for derivation of these results. The formalism, and its demonstrations in novel experiments, appears to represent a new model for the measurement procedure. The simplicity of the result (\ref{delk}), and the insight it provides, can make it a valuable tool for devising high precision measurement protocols and for studying fundamental quantum concepts.

This work has been supported by the Israel Science Foundation  Grant No. 1125/10 and by the ERC DM Grant No 32103.

\bibliographystyle{elsarticle-num} 

\end{document}